\documentclass[aps,prb,showpacs,showkeys,nofootinbib,reprint,
superscriptaddress,preprintnumbers]{revtex4-1}
\usepackage{amsmath,amssymb,bm,mathrsfs}
\usepackage{srcltx}
\usepackage{dsfont}
\usepackage{epsfig}
\usepackage{slashed}
\usepackage{bbold}
\usepackage{psfrag}
\usepackage{xcolor}
\PassOptionsToPackage{caption=false}{subfig}
\usepackage{subfig}
\usepackage{xfrac}
\usepackage{multirow}
\usepackage{booktabs}
\usepackage{hyperref}
\hypersetup{colorlinks=true,linkcolor=blue,citecolor=blue,urlcolor=violet}
\usepackage{relsize}
\usepackage{verbatim}
\usepackage{natbib}
\usepackage{float}
\usepackage{graphicx}

\begin{document}

\author{Subhadip Jana}
\email{subhadip.j@iopb.res.in}
\affiliation{Institute of Physics, Sachivalaya Marg, Bhubaneswar–751005, India}%
\affiliation{Homi Bhabha National Institute, AnushaktiNagar, Mumbai–400085, India}%

\author{Shwetha G.Bhat}
\affiliation{Department of Physics, Indian Institute of Science Bangalore–560012, India}%


\author{B.C.Behera}
\affiliation{Institute of Physics, Sachivalaya Marg, Bhubaneswar–751005, India}%


\author{L.Patra}
\affiliation{Condensed Matter Theory and Computational Lab, Department of Physics, Indian Institute of Technology, Madras, India}%

\author{P.S.Anil Kumar}
\affiliation{Department of Physics, Indian Institute of Science Bangalore–560012, India}%
\author{B.R.K.Nanda}
\affiliation{Condensed Matter Theory and Computational Lab, Department of Physics, Indian Institute of Technology, Madras, India}%

\author{D.Samal}
\affiliation{Institute of Physics, Sachivalaya Marg, Bhubaneswar–751005, India}%
\affiliation{Homi Bhabha National Institute, AnushaktiNagar, Mumbai–400085, India}%



\title{Evidence for  weak antilocalization-weak localization crossover and metal-insulator transition in CaCu$_{3}$Ru$_{4}$O$_{12}$ thin films}

\begin{abstract}
	Artificial confinement of electrons by tailoring the layer thickness has turned out to be a powerful tool to harness control over competing phases in nano-layers of complex oxides. We investigate the effect of dimensionality  on transport properties of $d$-electron based heavy-fermion metal CaCu$_{3}$Ru$_{4}$O$_{12}$. Transport behavior evolves from metallic to localized regime upon reducing thickness and a metal insulator transition is observed below 3 nm film thickness for which sheet resistance crosses $h/e^{2} \sim 25~$k$\Omega$, the quantum resistance in 2D. Magnetotransport  study reveals a strong interplay between inelastic and spin-orbit scattering lengths upon reducing thickness,  which results in  weak antilocalization (WAL) to weak localization (WL) crossover in magnetoconductance.
\end{abstract}
\maketitle
\section{Introduction}
Reducing the dimensionality of a system often engenders electromagnetic properties sharply different from their bulk counterpart. It mainly arises due to enhanced quantum effects and increased correlations due to reduction in available phase space and screening\cite{imada1998metal}. The interplay between  electron band width ($W$) and onsite Coulomb energy ($U$) in correlated electron system sensitively depends on the lattice  dimensionality of the electron system. Manipulation of  correlated electronic states in artificial crystal structures by exploiting their layer thickness down to unit cell level and epitaxial strain (without resorting to any chemical substitution that might induce unintentional disorder) is seen as a viable route to obtain more insight into these materials and renders a perfect platform to search for unforeseeable complex phenomena\cite{scherwitzl2011metal,yoshimatsu2010dimensional,boris2011dimensionality,king2014atomic,stemmer2018non,samal2013experimental}. Besides $U$  and $W$, spin-orbit interaction plays a significant  role in governing the underlying  electronic properties as evidenced in the study of Ir based oxides\cite{schutz2017dimensionality,gruenewald2014compressive,kim2008novel,matsuno2015engineering,rau2016spin}.
	
Conducting systems with strong  spin-orbit coupling (SOC)    often manifest weak antilocalization (WAL) effect and has extensively been explored in materials containing heavy elements  (like Bi,Ir,Pt,Au)\cite{matetskiy2018weak,jenderka2013mott,beckmann1996first}. Besides,  Dresselhaus/ Rashba type SOC effects in systems that  lack  bulk inversion symmetry/asymmetry in confining potential   (for e.g.  modulation doped semiconductor hetero-structure GaAs/Al$_{x}$Ga$_{1-x}$As, and LaAlO$_{3}$/SrTiO$_{3}$) has triggered diverse research\cite{knap1996weak,stornaiuolo2014weak,chen2010gate,koga2002rashba}.  It is observed that applying an electric field across interface in the above cases induces WAL-WL  crossover. In present study, we demonstrate a possibility for WAL-WL crossover by systematically manipulating layer thickness in heavy fermionic CaCu$_{3}$Ru$_{4}$O$_{12}$ (CCRO) system with cubic   symmetry  (space group Im-3).
	
Layered ruthenates with different dimensionality, so called nature's engineered Ruddlesden-Popper (RP) type phases have attracted significant attention since they exhibit unique electronic and magnetic ground states such as unconventional spin-triplet superconductivity, metamagnetism and electron nematic phase, highly conducting ferromagnetism and spin glass behaviour\cite{burganov2016strain,stemmer2018non,cao1999antiferromagnetic,qu2009complex}.  The interplay among electron correlation, spin orbit interaction and dimensionality makes ruthenates more promising and a small perturbation can readily the tip the balance and promote unexpected changes in the electronic property. CCRO is intriguingly debated as a rare class of $d$ electron based heavy-fermion metal with excellent metallic conductivity along with a signature for broad  hump in magnetic susceptibility around 150-200K\cite{kao2017origin,hollmann2013correlation,krimmel2008non,tran2006electronic}. If the heavy fermionic metals can be made 2D,  unprecedented quantum phenomena are expected to result, and such studies are very much desirable.

\begin{figure}[H]
	\includegraphics[width=8.5cm ,height=3.5cm]{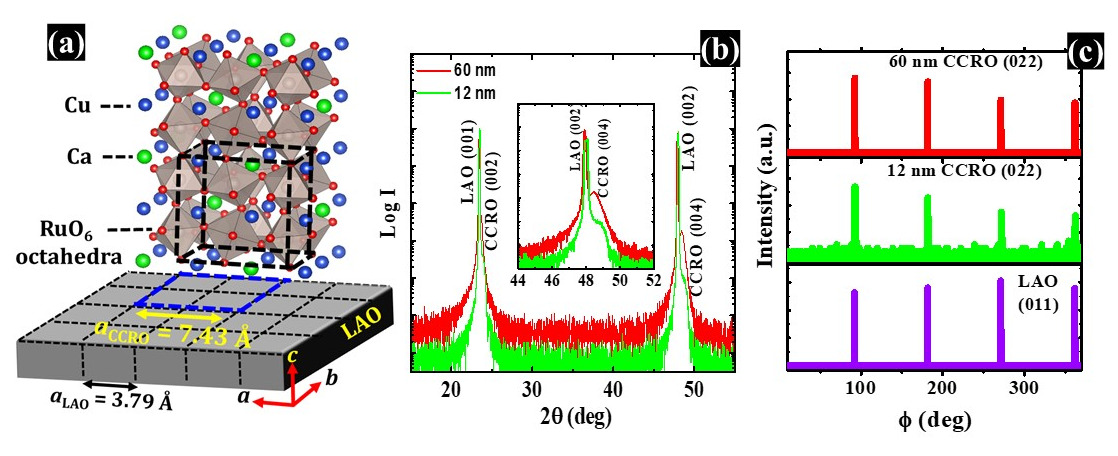}
	\caption{{(a)Schematic structure of CCRO on LAO. (b) $2\theta-\theta$ XRD pattern for (\textit{001}) oriented CCRO film on LAO(\textit{001}) of 12 nm and 60 nm thick films. (c) $\varphi$ scan about(\textit{011}) of LAO and (\textit{022}) of CCRO films (60 nm and 12 nm) showing  the epitaxial relation.}
	\label{fig:xrd}}
\end{figure}
We fabricate single-crystalline epitaxial CCRO thin films of varying layer thickness and investigate the dimensional effect on its  magnetotransport properties. We observe that upon reduction of CCRO thickness, transport behavior evolves from metallic to localized regime and a thickness driven metal insulator transition (MIT) is observed below 3 nm  for which the room temperature sheet resistance crosses $h/e^{2} \sim 25~$k$\Omega$, the quantum resistance in 2D\cite{licciardello1975constancy}. More importantly, from magnetoconductance we observe a strong interplay among  inelastic ($L_{Th}$) and spin-orbit ($l_{so}$ ) scattering lengths that gives rise to weak antilocalization (WAL) - weak localization (WL) crossover upon reducing thickness. Using 2D magnetotransport theory and  magnetotransport measurement, we elucidate the evolution of different types of scattering process (i.e. spin-orbit, phase breaking)  with variation of thickness.
	
\section{Thin Film Growth and Structural Characterization}
A series of high-quality single-crystalline epitaxial CCRO films with varying thickness were grown on LaAlO$_{3}$ (001) substrates ($a =3.79$ \AA) using pulsed laser deposition (PLD) with a KrF excimer laser ($\lambda = 248~$nm). The polycrystalline CCRO PLD-target material was prepared by stoichiometrically mixing CaCO$_{3}$, CuO, RuO$_{2}$ and heating the mixture for 26 hours at 1050$~^{0}$C temperature in open air and under ambient pressure \cite{krimmel2008non}. Single phase was obtained after many cycles of heating and grinding. During thin film growth, the substrate temperature and the O$_{2}$ partial pressure was maintained at 650$~^{0}$C and $5\times 10^{-3}~$mbar respectively and the deposition was carried out with a laser fluence $\sim$3 J/cm$^{2}$. Before the deposition, target was pre-ablated for 2 minutes to remove any possible surface contamination. The structural details of the films were characterized by X-ray diffraction (XRD) scan (using Rigaku Smart-Lab X-ray diffractometer in parallel-beam geometry with high resolution of Cu-K$_{\alpha}~$ radiation). The thicknesses of the films were calibrated with the number of laser pulses using cross-sectional SEM. Bulk CCRO exhibits cubic symmetry (Im-3 space group (No: 204)) with a lattice parameter of 7.43 \AA \cite{krimmel2008non} which is close to twice the lattice parameter of LaAlO$_{3}$ (LAO) (Fig.~\ref{fig:xrd}(a)). Thus, one can expect an in-plane tensile strain of $+1.97$\% for CCRO films with a cube-on-cube epitaxy on LAO. Fig.~\ref{fig:xrd}(b) exhibits wide-angle diffraction pattern for representative 60 nm and 12 nm thickness of CCRO films, indicating a c-axis oriented growth. The inset to Fig.~\ref{fig:xrd}(b) shows a zoom-in view of diffraction peak around \textit{(002)} of LAO. It is observed that the CCRO peak is close to the substrate for 60 nm than the 12 nm, signaling a compressed out-of-plane lattice in latter case. To verify the in-plane epitaxial relationship in these films, the $\varphi$ scans were performed on both 60 nm and 12 nm films about \textit{(022)} plane of CCRO in the vicinity of \textit{(011)} of LAO (Fig.~\ref{fig:xrd}(c)). Four equally spaced distinct peaks with a relative separation of $90^{0}$ (four-fold symmetry) were observed, suggesting an epitaxial growth of CCRO layer on LAO i.e.: [\textit{100}] LAO$\parallel$ [\textit{100}] CCRO. Altogether, our structural characterization implies an epitaxial \textit{(001)} oriented growth of CCRO on LAO \textit{(001)} substrate.  

\section{Electron Transport}
\begin{figure}[H]
	\includegraphics[width=4cm ,height=6cm]{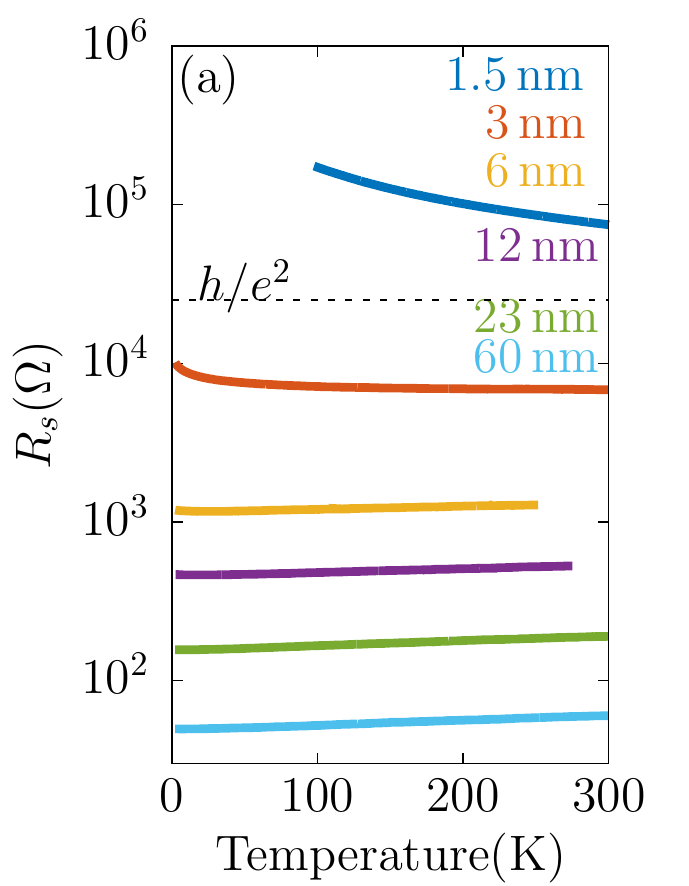}
	\includegraphics[width=4cm ,height=6cm]{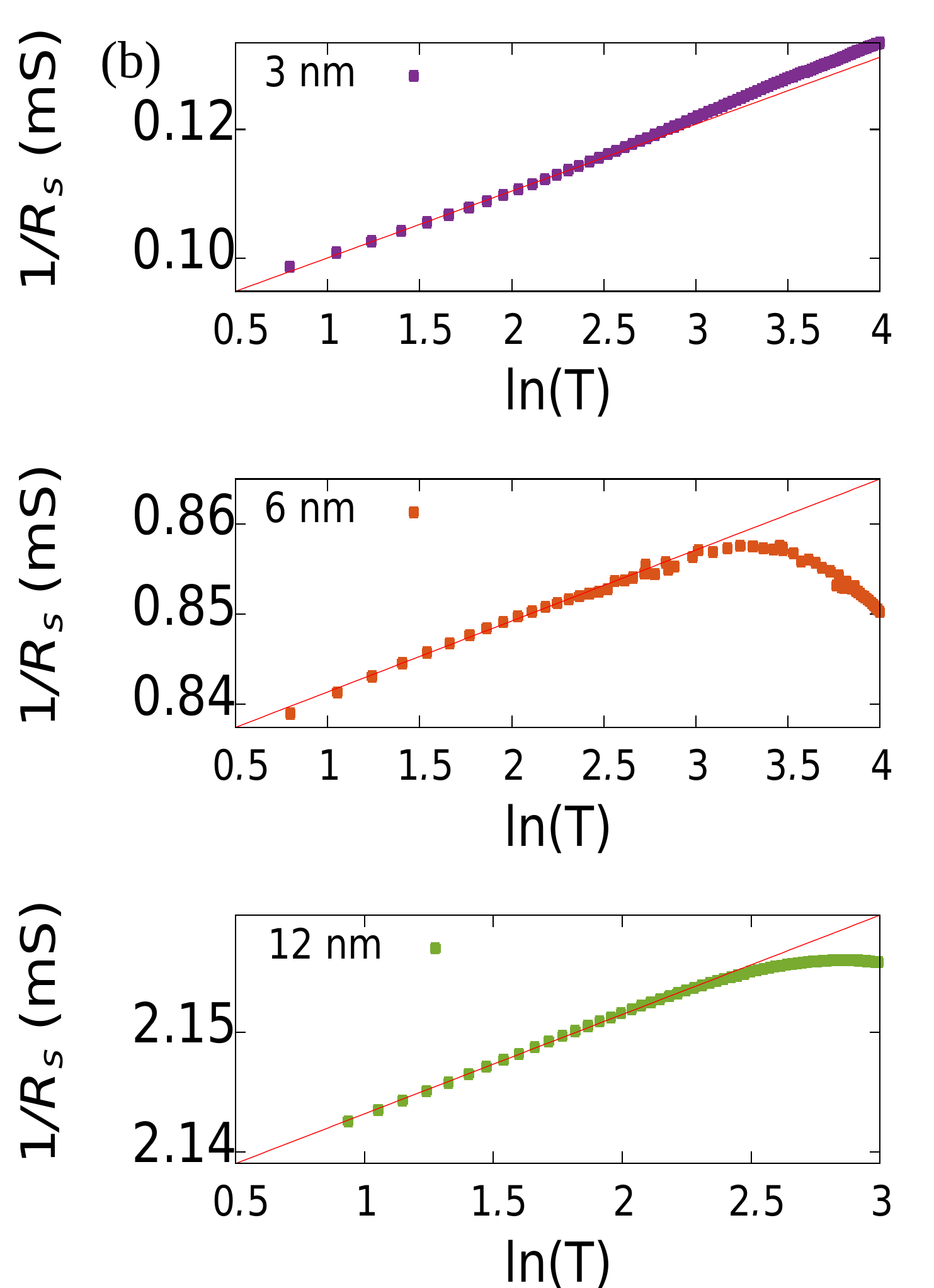}
	\caption{{(a) $R_{S}(T)$ for films of different thickness (MIT using Ioffe-Regal criterion is denoted by dotted line), (b) $1/R_{s}~vs~\ln(T)~ $  plots for films of  various thickness that reveals  the signature for quantum interference effect  in the  low  temperature regime.}
	\label{fig:rt}}
\end{figure}

We investigate the electrical transport properties of CCRO films with thicknesses ($t$) ranging from 1.5 to 60 nm.  Fig.~\ref{fig:rt}(a) shows the variation of sheet resistance ($R_{S}$) as a function of temperature. For films with t $\geq$ 3 nm, $R_{s}$ values are found to be below 25 k$\Omega$~ $(h/e^{2}  \simeq 25 ~$k$ \Omega)$ across the whole temperature range (2$-$300 K). However, the value of $R_{s}$ for the thinnest film with $t=1.5$ nm crosses 25 k$\Omega$ indicating a transition to insulating state. Based on Ioffe and Regel criterion,  MIT is expected in the limit of $k_{F}l_{e} \simeq 1~$  where  $k_{F}, ~l_{e}$  are Fermi wave vector and  mean free path respectively \cite{ioffe1960non}  and an approximate value for the Mott-Ioffe-Regel limit in two dimensions is $h/e^{2}~~$ \cite{licciardello1975constancy}. The systematic increase in $R_{s}$ and its rise above Ioffe-Regel limit indicates the occurrence of localization effect as the thickness of CCRO film is reduced.
\subsection{Arrhenius and Variable Range Hopping behaviour}
\begin{figure}[H]
	\begin{center}
	\includegraphics[width=4.5cm ,height=3.5 cm]{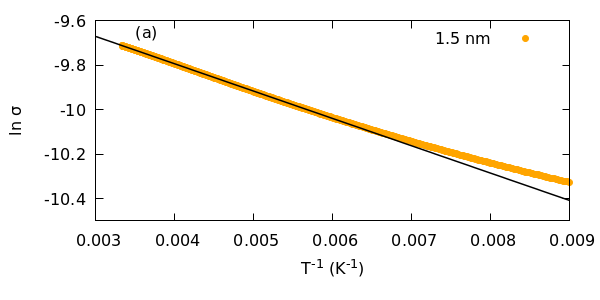}
	\includegraphics[width=4.5cm ,height=3.5 cm]{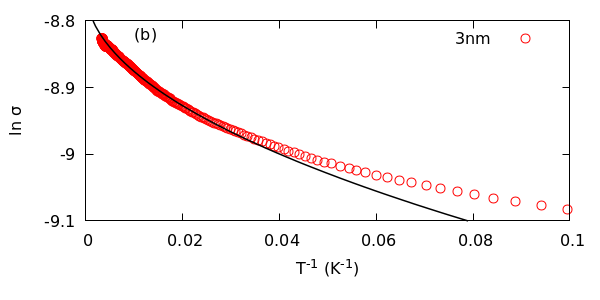}
	\caption{{(a) Resistivity  fitting  with  Arrhenius  for  1.5 nm, (b)  VRH  (ES type) 3  nm thick  film respectively (black solid lines are theoretical expressions).  }
	\label{fig:vrh}}
	\end{center} 
\end{figure}

Transport in 1.5 nm  film follows Arrhenius type behavior  $\sigma\propto\exp(-E_{g}/2kT)$  and  yields  an activation energy gap $E_{g}$ of 31 meV,  which is obtained  by fitting the data in the temperature range 300-160 K (Fig.~\ref{fig:vrh}(a)). On the contrary for the case of  3 nm film, which exhibits a negative temperature coefficient ($d\rho/dT < 0$) and is on the verge  MIT, the transport behavior can be described by a variable range hopping (VRH) type of conduction. VRH conduction scenario involves hopping of electrons between the localized electronic states within narrow bands close to the Fermi energy and conductivity $\sigma$ is given by $\sigma = C \exp[-(T_{0} /T)^{\alpha}]~$, 
where $T_{0}$ depend on the density of localized states and spread of wave functions. VRH conductivity can be either of  Mott or Efros-Shklovskii (ES) type and for a two dimensional case they    are characterized by the exponent $\alpha=1/3$  and  $\alpha=1/2$  respectively \cite{efros1975coulomb,mott1969conduction}. In our case for 3 nm CCRO film, the fit to the  conductivity data in the temperature range 50 K $\leq$T$\leq$ 300 K (Fig.~\ref{fig:vrh}(b)) yields $\alpha = 0.501$ (ES type)   indicating  the existence of  Coulomb   charge gap.
A closer inspection of $~R_{s} ~vs~ T~$  plot on the low temperature side reveals upturn in the metallic regime (not shown here)  and this could be attributed to quantum interference effect.  In presence of quantum interference effect conductance shows logarithmic temperature dependence   which can be expressed as (for 2D), \cite{lee1985disordered}\\
$\sigma(T) = \sigma_{0} + \frac{p e^{2}}{2 \hbar \pi^{2} }\ln\left(\frac{T}{T_{0}}\right)$,
where $\sigma(T)$ is sheet conductance at temperature $T$,  $~~p$ is defined as  $~L_{Th} = aT^{-p/2}~ $ and $T_{0}~$ is constant. \\
Fig.~\ref{fig:rt}(b) shows the logarithmic temperature dependence of reciprocal sheet resistance for representative films (3, 6, 12 nm)  which is fitted well in low temperature regime.
	
\section{Magnetotransport}
To examine the changes in  electron scattering process with the variation of film thickness, magnetotransport measurements were performed using physical property measurement system (PPMS Quantum Design) in four probe geometry, down to 2 Kelvin temperature and a magnetic field upto 8 Tesla.

\begin{figure}[H]
	\includegraphics[width=8cm ,height=6.6 cm]{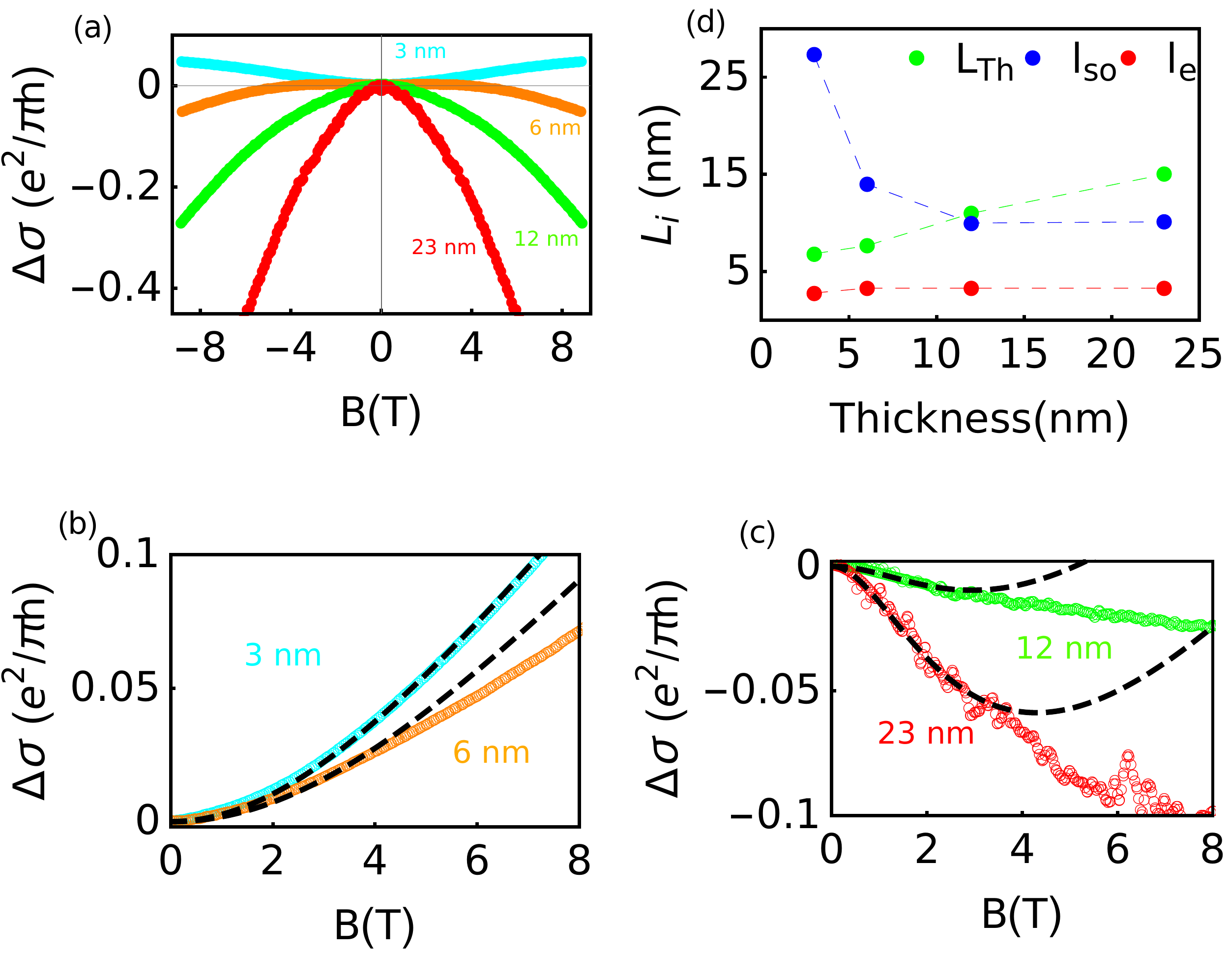}
	\caption{\label{fig:mc}{(a) Magnetoconductance$~\Delta\sigma=\sigma(B)-\sigma(0)$ measured for films at 2 K temperature including classical contribution ($\propto B^{2}$). (b),(c)Fitted with HLN equation (dashed black lines) after subtracting $B^{2}$ contribution from experimental data, \cite{jenderka2013mott}. (d) $L_{Th},l_{so},l_{e}$ extracted from fits.}}
\end{figure}

\textcolor{black}{Quantum interference among scattered electrons depends on their trajectory configuration. The interference correction to classical Drude conductivity tend to vanish in most cases after averaging over random scattering centers, except for the scattered electrons which propagate in identical time-reversed closed trajectories that gives rise to quantum interference correction to conductivity. Such kind of trajectories are known as ``Cooperon loop\textquotedblright(CL)  and depending upon constructive (destructive) interference among the scattered electrons in CL, it manifests WL (WAL) effect.}
These quantum effects can be captured by measuring conductance in presence of magnetic field. Fig.~\ref{fig:mc}(a) shows the measured  out of plane magnetoconductance (MC)   $\Delta\sigma$ in units of $~\frac{e^{2}}{\pi h}~$ at 2 K.  Fig.~\ref{fig:mc}(b),(c) shows the same after subtracting classical $B^{2}$, contribution that  arises due to Lorentz force  for film thicknesses ranging from 23 to 3 nm. A negative MC is observed for films with larger thickness i.e  t $\geq$ 12 nm. However, a crossover from negative to positive MC occurs as we reduce thickness. We attribute the positive (negative) magnetoconductance to WL (WAL) effect.
	
To  shed light on the observed magnetoconductance crossover, we fit the magnetoconductance curves with Hikami-Larkin-Nagaoka (HLN) equation \cite{hikami1980spin}  in 2D limit to extract the characteristic scattering lengths. For the convenience of experimental data fitting, we  express extended  form of HLN equation (Eq.\ref{eq1}) in 2D limit in terms of elastic, inelastic and spin-orbit scattering lengths which are denoted as $~l_{e},~L_{Th},~l_{so}~$respectively (see Supplementary Material for detailed derivation).
\begin{equation}\label{eq1}
 \begin{split}
	\Delta\sigma(B) &=   - \dfrac{e^{2}}{2\pi^{2}\hbar} \bigg[  {\psi(1/2 + B_{e}/B) + \log(B/B_{e}) } \\
	& ~ +\frac{1}{2}\Big\{\psi(1/2 + B_{Th}/B) + \log(B/B_{Th})\Big\} \\ 
	& ~-\frac{3}{2} \Big\{\psi(1/2 + \frac{B_{Th} + B_{so}}{B}) + \log(\frac{B}{B_{Th} + B_{so}})\Big\} \bigg] 
	\end{split}
\end{equation}
where \textcolor{black}{$\psi(x)$ is digamma function,} $B$ is external magnetic field and $ B_{e}, B_{Th}, B_{so}$ are related to $l_{e}, L_{Th}, l_{so}$  with following relations $B_{i} = \hbar /4el^{2}_{i}$ respectively. 2D limit is defined as  film    thickness ($t$) which is less than $L_{Th}$  (where $l_{i}=(D\tau_{i})^{1/2}$); where $D$ is  Diffusion coefficient and  $D \propto v^{2}_{F}$ ($v^{}_{F}$ is Fermi velocity) and $\tau_{i}$ is  scattering time for corresponding length.
\textcolor{black}{After subtraction of classical contribution to magnetoconductance $\propto B^{2}~~$ \cite{jenderka2013mott}}, 
$\Delta\sigma(B)~$ is  fitted  as per Eq.\eqref{eq1}    which is shown in  Fig.~\ref{fig:mc}(b),(c).   The extracted lengths from the fitting are shown in Fig.~\ref{fig:mc}(d). The extracted lengths show a crossover from $L_{Th} <  l_{so}$ for the films with thickness  3, 6 nm and   $L_{Th} > l_{so}$ for  12, 23 nm which is consistent with WL-WAL crossover.

Below we discuss case by case study for the observed magneto conductance crossover with thickness variation. It is evident from Fig.~\ref{fig:mc}(b) that 3, 6 nm thick films show  WL effect (i.e $\Delta\sigma(B)$  increases with increment of magnetic field). Since $L_{Th} < l_{so}$ (for 3, 6 nm films)  the spin-orbit related scattering effect becomes redundant (spin-orbit interaction strength $ \propto 1/l^{2}_{so}$) for the phase space in which quantum coherence is maintained, electron encountering SO scattering is very weak. In other words, small $L_{Th}$ allows CL in very short length scale where SO interaction is not able to rotate the spin and thus results in WL due to constructive interference.  In WL regime, magnetic flux induces additional phase difference between two electrons  moving in time-reversed identical closed trajectory and this additional phase difference breaks constructive interference and as a result conductivity increase with the application of magnetic field.  For films with t $\geq$ 12 nm, $L_{Th}$ tends to increase and $l_{so}$ gradually decreases. Therefore electrons are able to move relatively more distance by maintaining phase coherence and simultaneously due to smaller value of $l_{so}$, electrons encounter spin-orbit elastic scattering within a shorter distance. Therefore  WAL becomes prominent with increasing thickness and is a cumulative effect of above two factors.
\section{Conclusion}
	
In conclusion, we have demonstrated that epitaxial single crystalline thin films of heavy-fermion correlated metal CCRO grown on LAO can be driven to an insulating state  upon reducing its thickness down to 1.5 nm.  More importantly, we find that magnetoconductance shows a crossover from WAL to WL behaviour  with reduction of film thickness. From  analysis of magnetotransport data, we realize that   a subtle interplay between effective spin-orbit  scattering and inelastic scattering of electrons could render such crossover. Our study  elucidates an important  role of dimensionality on  quantum transport behavior in  CCRO.

	
\begin{center}
	\bf{ACKNOWLEDGEMENT}
\end{center}

S.J and D.S acknowledge  V.Tripathy, S.Mandal, D.S and B.C.B  acknowledge  the  financial  support  from  Max-Planck Partner Group. B.R.K.N acknowledges Department of Science and Technology, India, for Grant  No-EMR/2016/003791. S.G.B  would like to acknowledge  INSPIRE Faculty award, DST, INDIA for the financial support. P.S.A.K acknowledges Nano Mission, DST, INDIA for funding support.	
\bibliography{ref.bib}
\begin{figure}
	\includegraphics[page=1,scale=0.85]{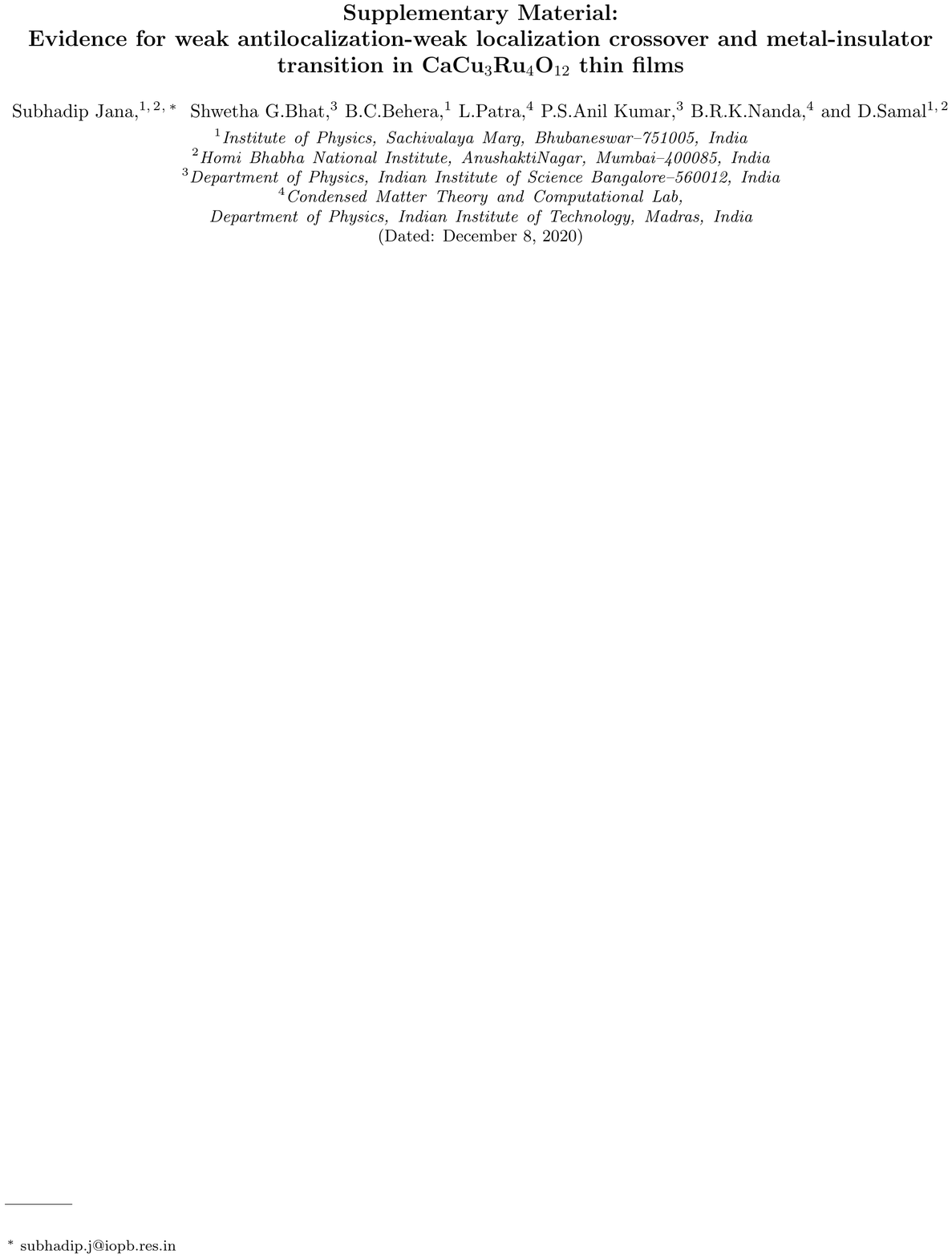}
\end{figure}

\begin{figure}
	\includegraphics[page=2,scale=0.85]{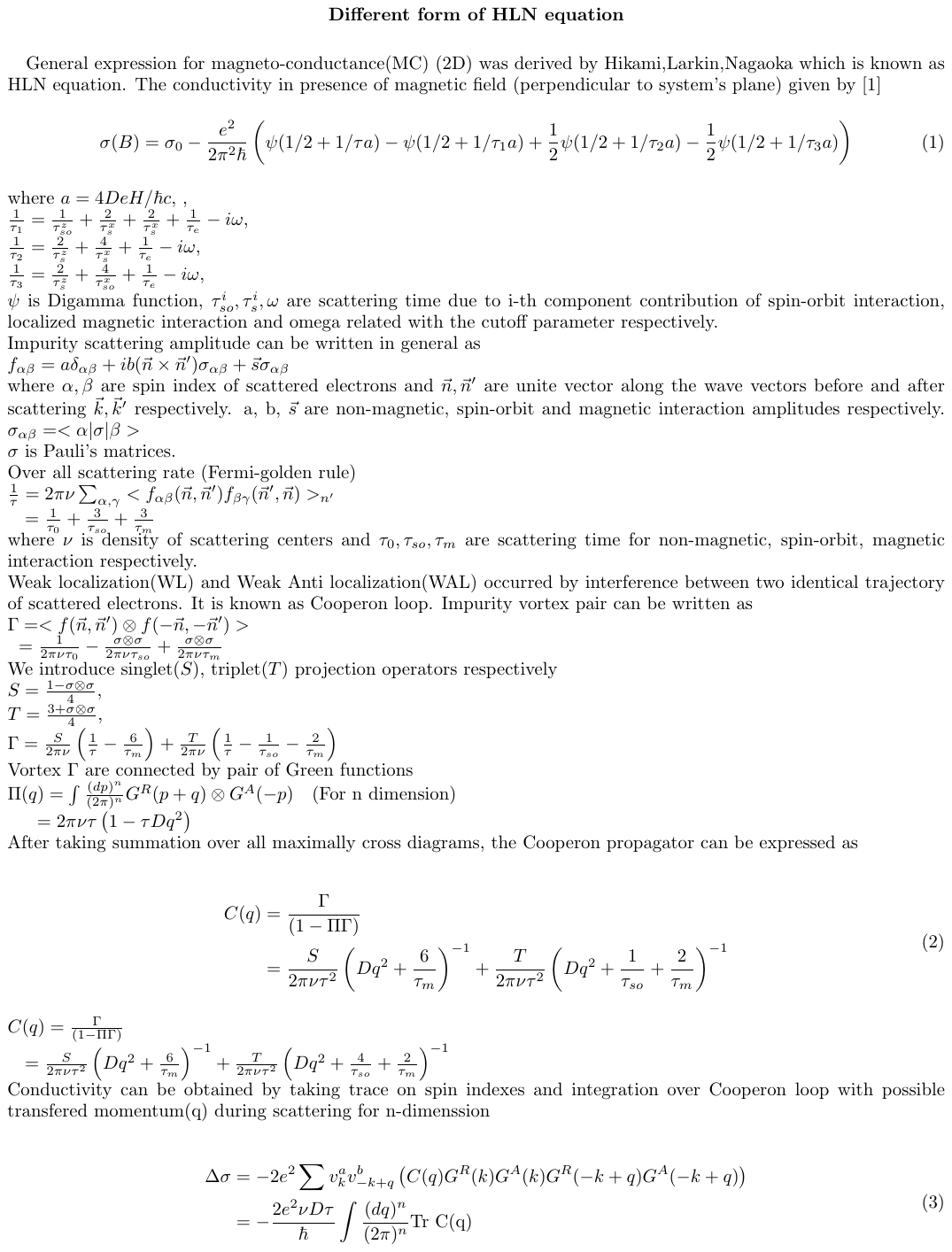}	
\end{figure}
\begin{figure}
	\includegraphics[page=3,scale=0.85]{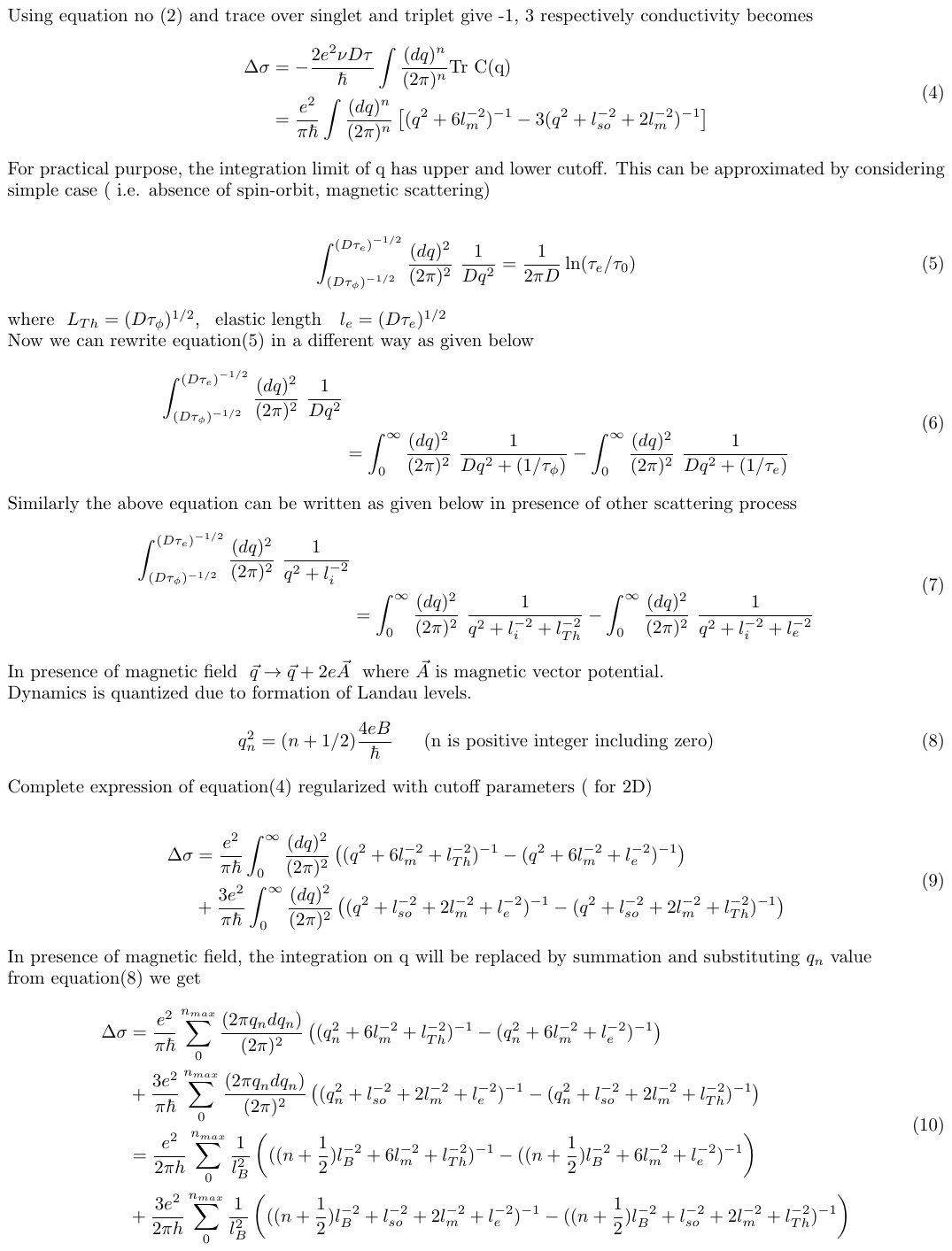}
\end{figure}
\begin{figure}
	\includegraphics[page=4,scale=0.85]{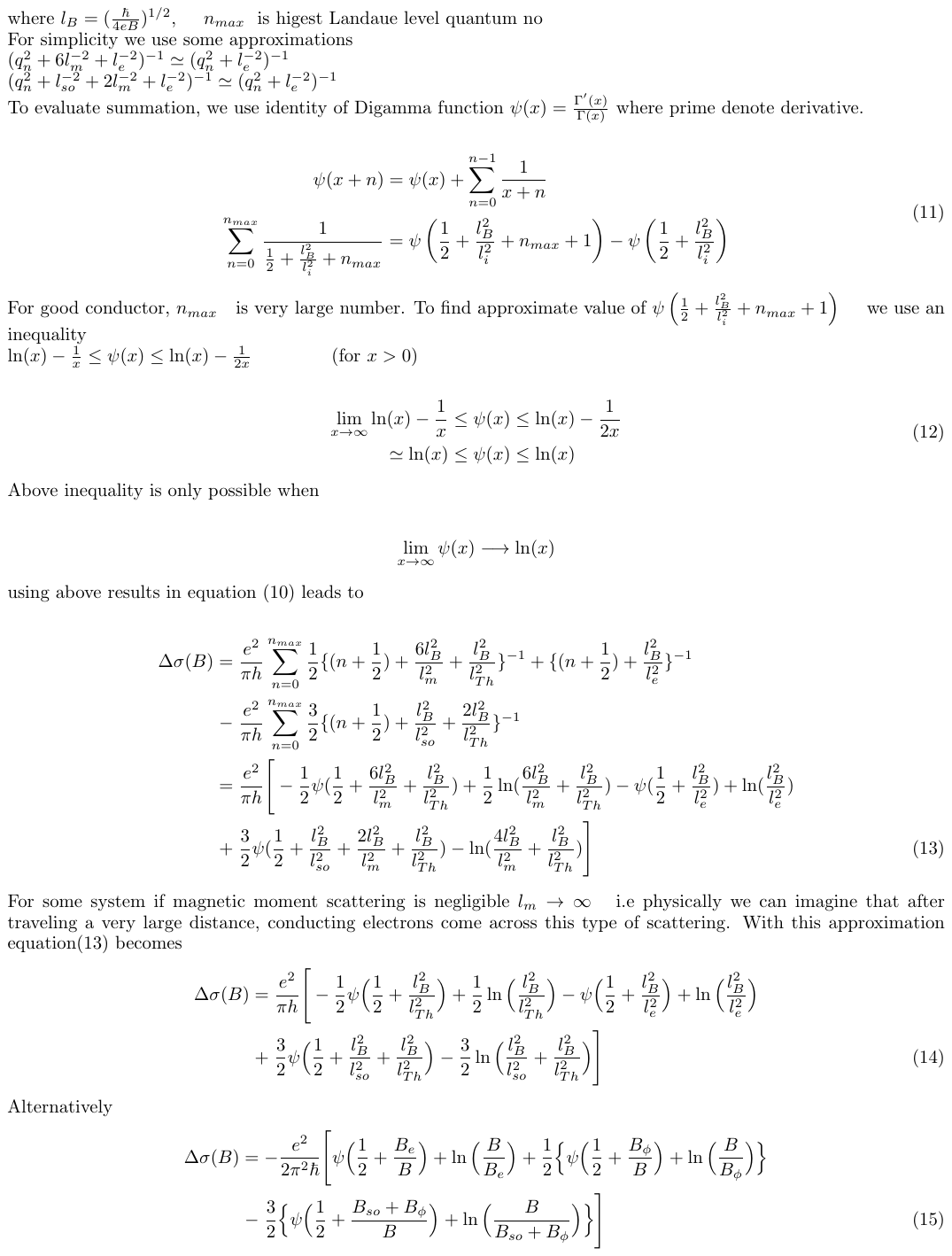}
\end{figure}
\begin{figure}
	\includegraphics[page=5,scale=0.85]{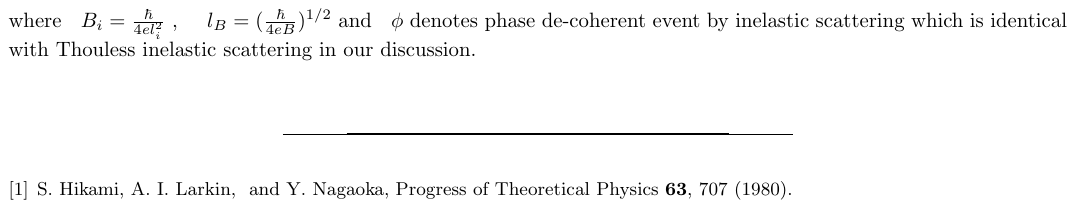}
\end{figure}

\end{document}